\documentclass[
    ,final            
  ]
  {aipproc}

\layoutstyle{6x9}

\begin{document}

\title{Chiral doublings of heavy-light hadrons:\\ 
New charmed mesons discovered by BABAR, \\CLEO and BELLE}

\author{Maciej A. Nowak }{
  address={M. Smoluchowski Institute of Physics, Jagiellonian University,
\\
30-059 Krak\'{o}w, Reymonta 4, Poland\\
e-mail: nowak@th.if.uj.edu.pl}
}

\begin{abstract} 
We remind the chiral doubling scenario\cite{us92,bh93} for hadrons built
of heavy {\em and} light quarks. Then we recall the  arguments why 
new states $D_s(2317)$, $D_s(2460)$, $D_0(2308)$ and $D_1^{'}(2427)$
should be viewed as chiral partners of $D_s$, $D_s^{*}$, $D$ and $D^*$, 
respectively. We summarize with the list of predictions based of 
chiral doubling scenario for other heavy-light
hadrons.   
\end{abstract}
\maketitle

\def\be{\begin{eqnarray}} 
\def\ee{\end{eqnarray}}
\newcommand{\no}{\nonumber}

Recently, experimental physics of hadrons with open charm 
has provided several spectacular discoveries:\\
-- First, BaBar~\cite{BB} has announced new, narrow meson
 $D^{*}_{sJ} (2317)^+$, 
decaying into $D_s^+$ and $\pi^0$. This observation was then confirmed by 
CLEO~\cite{CLEO}, which also noticed another narrow state,
 $D_{sJ}(2463)^+$, decaying
into $D_s^*$ and $\pi^0$. Both states were confirmed 
by Belle~\cite{BELLE}, and finally, 
the CLEO observation was also confirmed by BaBar~\cite{BABAR2}. 
From the moment of discovery, both states triggered a flurry of activity 
among the theorists. Experimental results were surprising, since such states 
were neither   expected below the $DK$ and $D*K$ thresholds 
nor were expected to be 
that narrow.
 Till today, several theoretical constructions were proposed 
to explain the masses, quantum numbers and decay patterns, most of them 
discussed during this conference.\\ 
-- Second, Belle has not only measured the narrow excited states 
$D_1,D_2$ with foreseen  quantum numbers $(1+,2+)$, 
but provided also first evidence
for two new, broad states $D_0*$ $(2308\pm 17\pm 15\pm 28)$
and $D_1^{'}$ $(2427\pm 26 \pm 20 \pm 17)$. Both of those  are {\em ca} 
$350-400$ MeV 
higher above the usual $D_0,D^{*}$ states and seem to have opposite parity.\\ 
-- Third, Selex  has provided preliminary data for 
   doubly charmed baryons~\cite{SELEX}. On top of known since December
   $ccd$ state $(3520)$, four other cascade $j=1/2$ states are visible. 
   Their masses are a challenge for   standard estimations based on potential models~\cite{HHL}.  This calls for alternative predictions, based either on 
  chiral solitonic models or diquark scenarios.

In this talk, we recall that actually the presence of this type of 
states was predicted by theoretical arguments already in 1992 and
1993, and is in fact required from the point of view of symmetries
of the QCD interactions. The two, mentioned above particles 
 observed by BaBar and
CLEO  are the first, theoretically anticipated~\cite{us92,bh93}
  {\em chiral partners}  of hadrons 
 built out of light and heavy quarks. As such, they should
represent rather a {\em pattern} of spontaneous breakdown of
chiral symmetry than isolated events.

Strong interactions involve three light flavors  (u, d, s) and
three heavy flavors (c, b, t) with respect to the QCD infrared
scale. The light sector (l) is characterized by the spontaneous
breaking of chiral symmetry, while the heavy sector (h) exhibits
heavy-quark (Isgur-Wise) symmetry \cite{IW}. In our original
work~\cite{us92} we addressed the question of the form of the
heavy-light effective action in the limit where light flavors are
massless, while the heavy flavors are infinitely massive.
The novel aspect of our original derivation was that consistency
with the general principles of spontaneously broken chiral
symmetry requires the introduction of {\em chiral partners} in the
form of a $(0^+,1^+)$ multiplet of pseudoscalars and
transverse vectors~\cite{us92}.
In the heavy-quark limit, the splitting between the chiral partners 
is small and of the order of the ``constituent
quark mass''.
The chiral corrections to the splitting 
were recently  shown to be of order $m_\pi^2/4m_h$, and therefore small
irrespective of an effective Lagrangian analysis~\cite{usnew}.

In brief, to leading order in the heavy-quark mass, the one-loop effective
action for the chiral doubler $(0^+,1^+)
$\be
G=
\frac{1+v\!\!\!/}{2}(\gamma^{\mu}\gamma_5\tilde{D}^{*}_{\mu} +
\tilde{D})\,\,.
\ee
 duplicates  the known action~\cite{wise}
for the standard $(0^-,1^-)$ multiplet
\be
H=
\frac{1+v\!\!\!/}{2}(\gamma^{\mu}D^{*}_{\mu} +
i\gamma_5D)\,\,.
\ee
i.e. 
 \be
{{\cal L}^G}=-\frac{i}{2} {\rm
  Tr}(\bar{G}v^{\mu}\partial_{\mu}G-v^{\mu}\partial_{\mu}
  \bar{G}G) + {\rm Tr} V_{\mu}\bar{G}G v^{\mu} - {\bf g}_G{\rm Tr}
   A_{\mu}\gamma^{\mu}\gamma_5\bar{G}G
 -{\bf m_G}(\Sigma)\,{\rm Tr}\bar{G}\,G
\label{notsousualcopy}
\ee
The axial $A_{\mu}$ and vector $V_{\mu}$ 
 currents are light currents, contributing to 
transitions  in odd or even number of pions, respectively 
(or generically, $SU(N_l)$ Goldstone bosons).
The key difference is the opposite sign 
in the sign of the constituent mass contribution in (\ref{notsousualcopy}),
with respect to similar term for $H$ multiplet. 
The sign flip follows
from the $\gamma_5$ difference in the definition of the
fields $H$ and $G$. 
 In other words:  it 
is sensitive to the parity content of the heavy-light field since
$H\rlap/{v}=-H$ and $G\rlap/{v}=+G$. The result is a split between
the heavy-light mesons of opposite chirality. This unusual
contribution of the chiral quark mass stems from the fact that it
tags to the {\em velocity} $H\rlap/{v}\bar{H}$ of the heavy field
and is therefore sensitive to {\em parity}.
The reparametrization invariance (invariance under velocity
shifts of the heavy quark to order one) introduces mass shifts that are
parity insensitive to leading order in $1/m_h$~\cite{NZ93}.
Chiral partners  communicate with  each other via light axial
currents
\be {{\cal L}_{HG}}=
\sqrt{\frac{{\bf g}_G}{{\bf g}_H}}\,
{\rm Tr}(\gamma_5\bar{G}H \gamma^{\mu} A_{\mu}) -
\sqrt{\frac{{\bf g}_H}{{\bf g}_G}}\,
{\rm Tr} (\gamma_5\bar{H}G\,\gamma^{\mu} A_{\mu})
\label{int}
 \ee
with no vector mixing because of the parity.

We visualize chiral doublers scheme for mesons in the form of cartoon, 
see  Fig.~1. 
\begin{figure}
\parbox{\textwidth}
{\centerline{
\includegraphics[width=.48\textwidth,height=.37\textwidth]{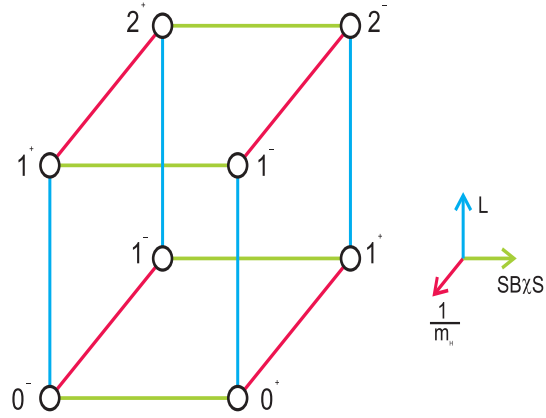}
}}
\caption{Cartoon representing {\em schematic} classification of chiral 
doublers.}
\end{figure}
 The three-dimensional ``cube'' is aligned along  three ``directions'':\\
- chiral symmetry breaking (horizontal, green)\\
- Isgur-Wise symmetry breaking (skew, red)\\
- total {\em light} angular momentum (vertical, blue).

The corners of the cube represent generic $h\bar{l}$ mesons, i.e. 
we expect similar ``cubic'' patterns  for 
$c\bar{s}$, $c\bar{u}$, $c\bar{d}$, $b\bar{s}$, $b\bar{u}$, $b\bar{d}$ mesons.   
Let us focus on $c\bar{s}$ states, i.e $D_s$-cube. 
Lower left rung represents
known pseudoscalar $0^-$  $D_s(1969)$ and vector $1^-$ $D_s^{*}(2112)$,
belonging to $j_l=1/2$ light angular momentum representation. 
The splitting between them is an $1/m_c$ effect and is expected to vanish 
in infinitely heavy charm quark limit, i.e. both particles would have form 
the $H$ multiplet. The upper left rung corresponds to $j_l=3/2$ representation,
i.e. $1^+$ and $2^+$ {\em excited} multiplet. Here $D_{s1}(2536)$ and 
$D_{sJ}^{*}(2573)$ are the candidates, separated by 
(smaller for excited states)
$1/m_c$ origin mass splitting. 
 Similar pattern applies for the 
non-strange charmed mesons (D-cube), i.e. $D(1865)$, $D^{*}(2010)$,
$D_1(2420)$ and $D_2(2460)$. This ``left plaquette'' 
 of the cube completes the standard, ``pre-BaBarian''
charmed meson spectroscopy. 

The novel aspect is the existence of the chiral doublers, i.e. 
the appearance of the {\em right plaquette}.  
First, we expect two chiral partners for $D_s$ and $D_s^*$, representing right
lower rung.
Here newly discovered $D^{*}_{sJ}(2317)$ and $D_{sJ}(2463)$ are 
the candidates for the $(0^+,1^+)$ scalar-axial $G$ multiplet. 
The averaged splitting for $(0^+,0^-)$ and the averaged splitting for 
$(1^+,1^-)$ are $349.2\pm 0.8$ and $346.8+1.1$, respectively, 
i.e. almost identical, as predicted a decade ago~\cite{us92,bh93}. 
Naturally, the splitting within the $G$ multiplet, i.e. between the masses
of the new BaBar state and CLEO state, is identical  to the 
splitting between the $(1^-,0^-)$ pair. 
The narrowness of the new states is basically the consequence of the 
kinematic constraints, as pointed by~\cite{beh}: since the chiral split 
is smaller than the mass of the kaon, these states live longer. 
On top of this effect, the isospin conservation most probably 
forces the pionic decay 
via  virtual $\eta$ decay, suppressing the rate even further~\cite{beh}.
Electromagnetic transitions, estimated on the basis of chirally
doubled lagrangians in~\cite{beh} are also in agreement with 
the experimental 
data. It is noteworthy to stress, that chiral Ward identities 
additionally constraint the amplitudes of the pionic decays 
for the $H$ multiplets,  $G$ {\em and} for  $G-H$
 pionic transitions~\cite{usnew}.

Let us move now towards the excited states. 
On the basis of the chiral doublers scenario, we would also expect 
the chiral partners for the excited $j_l=3/2$ multiplet, i.e. 
{\em new} chiral pair $(1^-,2^-)$~\cite{NZ93}. 
Alternatively, this pair could be 
also viewed as the $j_l=3/2$ excitation of the BaBar-Cleo $(0^+,1^+)$ 
multiplet.
 Our prediction for the masses of this {\em new} pair reads:
\be
m(\tilde{D}_{s1})&=&2721\pm 10 {\rm MeV} \nonumber\\
m(\tilde{D}_{s2})&=&2758\pm 10 {\rm MeV} 
\ee
where we used~\cite{usnew} as an
input the observed BaBar and CLEO splitting
for the chiral multiplet $(0^+,1^+)$ and the  mass formulae
obtained in~\cite{NZ93}. Note that the chiral splitting for excited 
states is approximately half of the chiral splitting for the ground pair. 

This completes the identification of corners of the ``cube''.
Left and right plaquettes are chiral copies, 
front and back plaquettes become degenerate in infinite mass of the 
heavy quark and the lower and upper plaquettes are separated by the 
 excitation of total {\em light} angular momentum $j_l$. 
We do not discuss here the possibility of even higher angular excitations, 
i.e. additional $j_l=5/2$ plateau, pointing only that first such states
$(2^-,3^-)$ may naturally appear above 3 GeV.

Let us move now towards non-strange charmed mesons. 
Here two states from Belle, $D_0^{*}(2308)$ and $D_1^{'}(2427)$ 
are natural candidates for  lower right rung of the D-cube, i.e.  
for the chiral doublers
of $D(1825)$ and $D^*(2010)$. There are however broad, since neither
kinematic nor isospin restrictions apply here, contrary to their
strange cousins. 

One of the arguments against the interpretation that the above pair 
might be  a chiral doubler is based on 
the values of  the  ``chiral'' mass shift, which 
(modulo experimental errors) seems to be equal of even larger 
 for the 
non-strange mesons than for the strange ones. 
Actually, this argument acts rather in favor 
for  the chiral doublers scheme~\cite{usnew}. 
A simple
parametrization for the constituent quark mass 
(with good comparison to instanton liquid model 
and lattice data)
was quoted in~\cite{mus}
 \be \Sigma (m_l) \approx m_l
+\Sigma(0) (\sqrt{1+(m_l/d)^2}-m_l/d) \label{massinst} 
\ee 
with
$\Sigma\approx 345\, {\rm MeV}/c^2$, $d\approx 198\,{\rm MeV}$.
 For a strange quark mass $m_s\approx
150\, {\rm MeV}$, the second term is reduced to $\Sigma(0)/2$,
making the combination (\ref{massinst}) weakly dependent on $m_l$
and of order $\Sigma$ all the way up to the strange quark mass.
Thus, both mass splittings are about the same for $(u,d,s)$
heavy-light mesons. Taking e.g. the value $\Sigma(0)=400$ MeV
one could easily estimate the effects of chiral splitting
for strange quark to be of order 350 MeV, so even smaller 
than for non-strange quarks.  
We would like to mention, that several simplified models 
just  add the {\em constant} chiral dressing to the current mass, 
ignoring the feedback of the explicit breaking of the chiral symmetry 
on the vacuum effects. 

Let us mention for completeness, that the chiral
doubling should be even more pronounced for bottom mesons, since
the $1/m_h$ corrections are three times smaller. For $m_s=150$
MeV, we expect~\cite{usnew} the chiral partners of $B_s$ and $B_s^*$ to be 323
MeV heavier, while the chiral partners of $B$ and $B^*$ to be 345
MeV heavier, i.e. close to predictions in~\cite{beh}. 
We note that any observation of chiral doubling for B
mesons would be a strong validation for chiral doublers  proposal. Indeed, in
the recently proposed alternative scenarios discussed during this conference
(multiquark states, hadronic molecules, modifications of quark
potential, unitarization) a repeating pattern from charm to bottom
calls for additional assumptions.

Finally, let us mention very briefly the consequences of the chiral doublers
scheme for the heavy light baryons, i.e. $hhl$ and $hll$ type. 
Already in~\cite{us92} we pointed, that the opposite parity baryons 
could be described as chiral solitons  of the action of 
({\ref{notsousualcopy}) type. In the light of presented here
preliminary Selex  data~\cite{SELEX} and planned COMPASS experiment, 
the issue of doubly heavy baryons is no longer academic. 
It is tempting to speculate, that the observed splitting
between the signals for the $ccu$ states of opposite parity, 
 3780 MeV and 3460 MeV~\cite{SELEX} is of chiral origin (here 340 MeV). 
Indeed, double heavy baryons, with the heavy diquark in spin 1 state
behaving alike the heavy color source $\bar{3}_c$ resemble mesonic
configurations and are subjected to chiral doubling due to the 
light quark physics, a point also noted recently by~\cite{beh}.   

We do not discuss here to 
what extent the newly discovered chiral partners can shed more 
light on effects of  
 charmonium absorption/regeneration in
thermal models with medium effects.
We stress that this is an important issue in
light of the current and future experiments at RHIC and  LHC as
well as at GSI, probing the restoration of the chiral symmetry. 

\begin{theacknowledgments}
This talk is based on work done in collaboration with Mannque Rho and Ismail 
Zahed~\cite{us92,usnew,NZ93}. This work was partially supported
by the Polish State Committee for Scientific Research (KBN) grant
2P03B 09622 (2002-2004). I am very grateful to David Cassel,
Murray Moinester, Jim Russ, Bob Cahn, Ken Hicks, Lonya Glozman, 
Pavel Krokovny, Kamal Seth,
Ted Barnes, Dan-Olof Riska, Su Houng Lee, Maxim Polyakov, 
Zhenya Kolomeitsev, Thorsten Feldmann,
Klaus Goeke, Frank Close and many others for passionate discussions
on new particles during this conference. 
\end{theacknowledgments}


{
}

\end{document}